# Observations of CMEs and Models of the Eruptive Corona

Nat Gopalswamy

*Code 671, NASA Goddard Space Flight Center, Greenbelt, MD 20771, USA*

**Abstract.** Current theoretical ideas on the internal structure of CMEs suggest that a flux rope is central to the CME structure, which has considerable observational support both from remote-sensing and in-situ observations. The flux-rope nature is also consistent with the post-eruption arcades with high-temperature plasmas and the charge states observed within CMEs arriving at Earth. The model involving magnetic loop expansion to explain CMEs without flux ropes is not viable because it contradicts CME kinematics and flare properties near the Sun. The flux rope is fast, it drives a shock, so the global picture of CMEs becomes complete if one includes the shock sheath to the CSHKP model.

## INTRODUCTION

The ejection of a magnetic structure in the form of a flux rope seems to be fundamental to the coronal mass ejection (CME) process. The flux rope has also been recognized as the basic structure in most numerical models of CMEs [1]. When the magnetic structure has an enhanced magnetic field strength, a smooth rotation in one of the transverse components, and low proton temperature, the structure is referred to as a magnetic cloud (MC) [2]. If one of the signatures is missing, the structure is referred to as non-MC. The non-MCs may be inherently lacking flux rope [3] or may have a flux rope not observed due to observational limitations [4]. The flux rope and non-flux rope models can be tested by comparing the flare structures at the Sun and the charge state distributions at 1 AU. The standard model of an eruption has been constructed based on various signatures such as H-alpha flare ribbons, post-eruption arcade rooted in the flare ribbons, and the eruption of a prominence overlying the polarity inversion line. Such a model is known as the Carmichael, Sturrock, Hirayama, Kopp, and Pneuman (CSHKP) model [5]. The eruptive filament is the basic mass motion signature in this model. When white-light CMEs were discovered, it was recognized that there are additional overlying structures that are spatially more extended than the prominence: an expanding coronal void region, and a bright frontal structure. The eruptive prominence has been found to be the core of CMEs observed in white light [6]. The void region seems to be the CME flux rope [7]. In CME literature, the frontal structure, void, and prominence core have been referred to as the "three-part structure". In most interplanetary CMEs (ICMEs), the leading structure is an interplanetary shock, followed by a compressed region of fluctuating magnetic field and enhanced temperature known as shock sheath and a magnetic structure that may or may not be a flux rope. A similar picture at the Sun has not been obvious for some time. Recent observations in white light and EUV do seem to indicate the presence of a shock, thus providing a picture similar to the in situ observations for fast CMEs. We present observational evidence for the flux rope model and the presence of shock signatures in the coronagraph images.

## DO ALL ICMES HAVE FLUX ROPE STRUCTURE?

Only about a third of CMEs observed in the IP medium appear as MCs, but this fraction varies over the solar cycle being the smallest during solar maxima and the largest during solar minima [8]. MCs and non-MCs have been explained by two different models: flare reconnection accompanied by flux rope formation for MCs and magnetic loop expansion for non-MCs. We discuss these two models and their observational consequences in this section.

### Flux Rope Model

The MC (or flux-rope) structure is thought to be formed by the reconnection process that also forms the post eruption arcade (PEA) as detailed in many papers [3,9,10]. This model calls for a very close relation between the flare and CME phenomena because each turn of the flux rope corresponds to an individual loop in the PEA. The azimuthal flux in MCs measured at 1 AU has been found to be similar to the reconnection magnetic flux in the flare ribbons, which strongly supports flux rope formation due to reconnection [11].

Thus, one can consider PEA as an indicator that a flux rope has been formed. One of the basic properties of PEAs is that they consist of hot plasma with temperatures much higher than the coronal temperatures, reaching tens of MK. The flux rope detaching from the PEA also contains this hot flare plasma. At high temperatures, heavy elements such as Fe and O are ionized multiple times, resulting in high charge states. As the flux rope expands rapidly after formation, the recombination becomes insignificant and the charge states remain unchanged in the flux rope. When the flux rope arrives at 1 AU, enhanced Fe and O charge states can be found in the ICME. Although initially no charge state enhancement was found in non-MC ICMEs [12], recent investigations find that both MC and non-MC ICMEs do have charge state enhancement [13, 14].

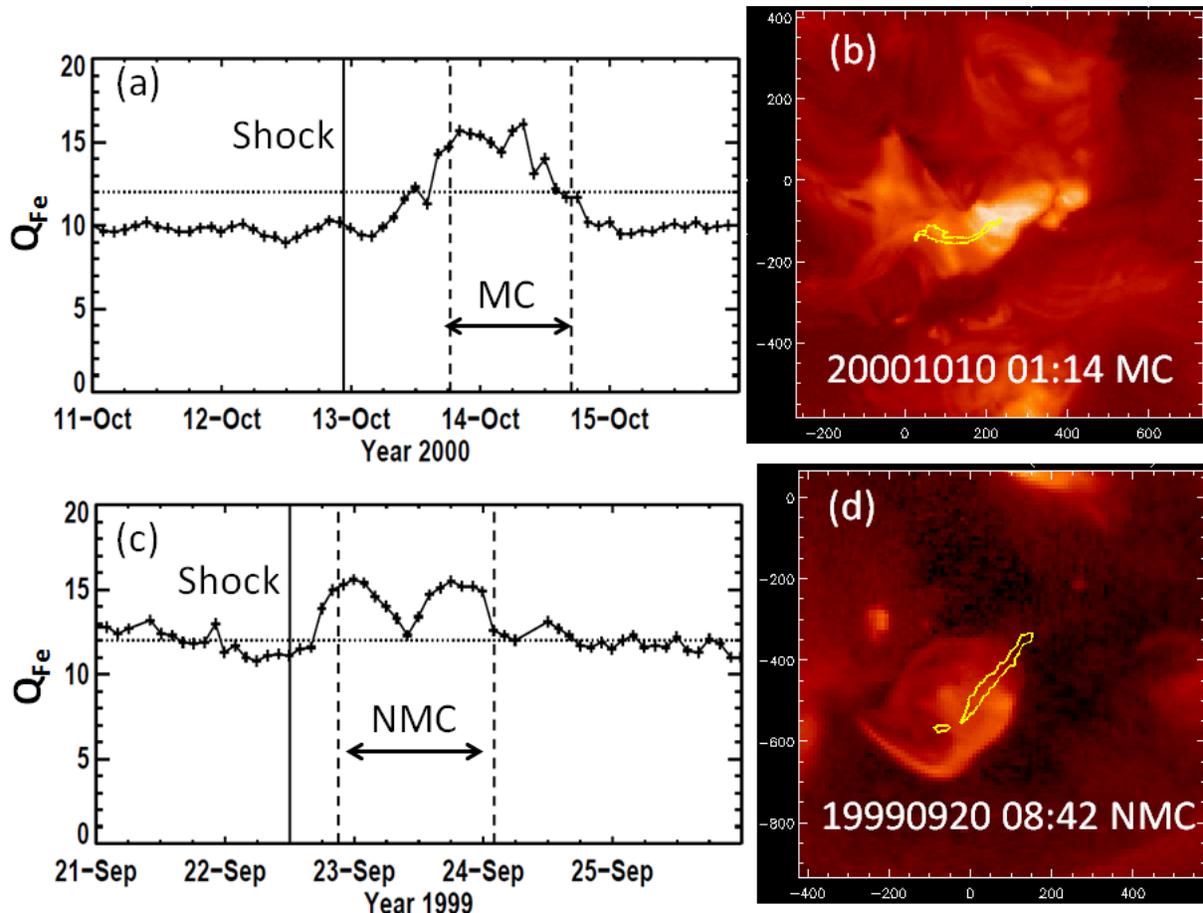

**FIGURE 1.** Average Fe charge states (a,c) measured by the Advanced Composition Explorer (ACE) mission and the post-eruption arcades in soft X-rays (b,d) for two ICMEs. The ICME on 2000 October 13 is a magnetic cloud, while that on 1999 September 22 is not a MC (NMC). The QFE value in the non-CME solar wind is typically less than 12 (the horizontal dotted line). The vertical dashed lines mark the ICME boundary obtained from solar wind plasma and magnetic signatures. The PEAs straddle the pre-eruption position of the prominence (outline superposed on the PEAs). .

Figure 1 compares the charge state enhancements in a MC and a non-MC and the solar source regions observed by the Yohkoh satellite's Soft X-ray Telescope (SXT). The average Fe charge state (QFe) is a measure of the charge state enhancement, determined by counting various Fe charge states at each instance. The typical value of QFe in the normal solar wind is ~11-12. Compared to the solar wind, both the MC and non-MC events show clear enhancements during the ICME interval. Both were shock-driving ICMEs, with no enhancement in the sheath region, except for a small interval in the sheath before the start of the ICME interval. The sheath enhancements can be attributed to the inaccurate definition of the start boundary of the ICMEs based on plasma and magnetic signatures. The main point is that QFe enhancement is similar in the MC and non-MC events. At the Sun, both events involved a filament eruption and PEA. Interestingly, the PEAs have similar morphology overlying the filament locations.

This means, both eruptions involved the formation of a flux rope and hot plasma was injected into the flux ropes. The case study presented here is consistent with the statistical study [14], which found essentially similar Fe and O charge state enhancements, and flare temperatures in a large number of MC and non-MC events originating from close to the disk center of the Sun. These results suggest that the flux rope formation process is common to both MC and non-MC events. In other words, both MCs and non-MCs have a flux rope.

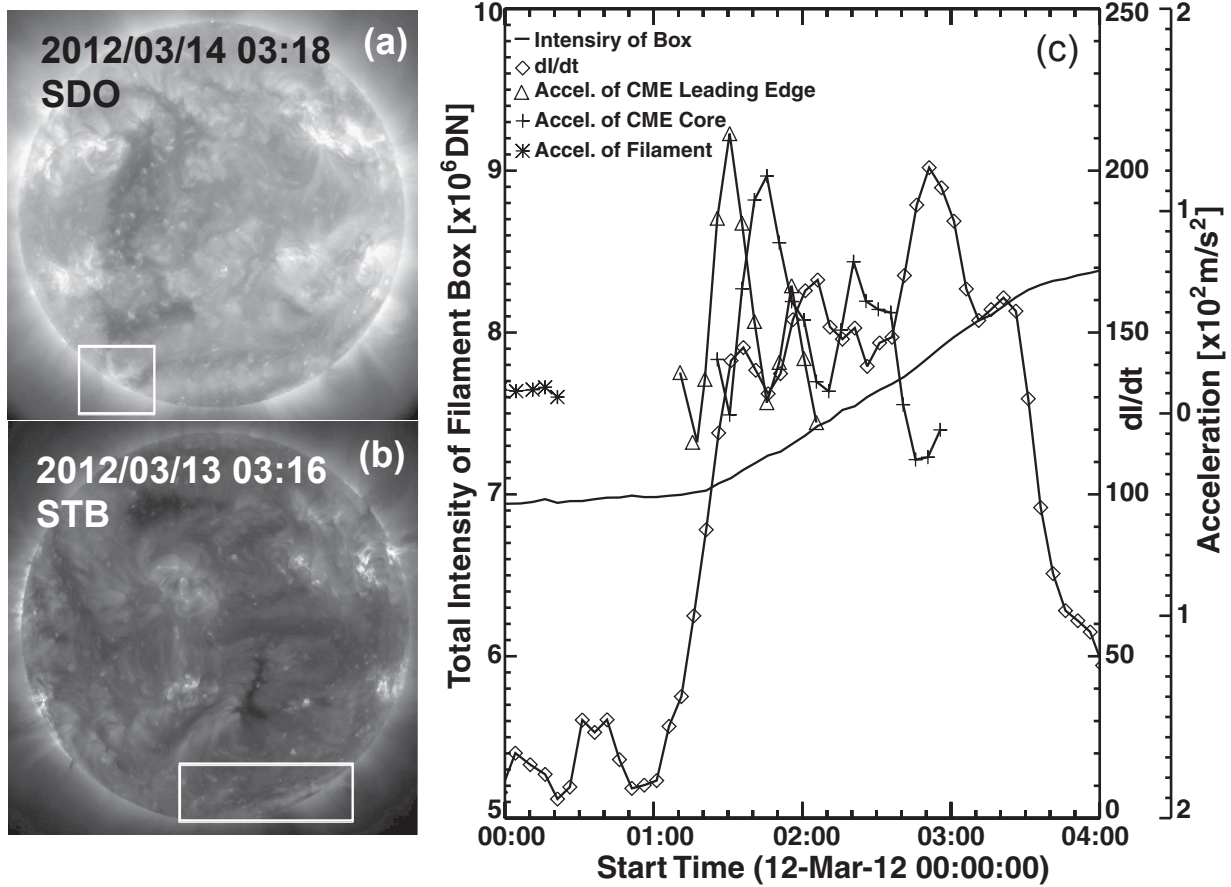

**FIGURE 1.** PEA observed in a PCF eruption observed by the Solar Dynamics Observatory (a) and the Extreme Ultraviolet Imager (EUVI) on board STEREO-B (b). The light curve of the PEA obtained within a box (I in Data Number units) around the PCF in (b) is shown in (c) along with its time derivative (dI/dt) and the measured accelerations of the CME leading edge and filament core observed in the STEREO's inner coronagraph (COR1) field of view. The situation is similar to the Neupert effect observed for regular flares.

## Loop Expansion Model

Another explanation for non-MC structures is that a simple loop expansion happens from closed field regions on the Sun resulting in interplanetary structures in which no smooth rotation is expected [3]. There is observational evidence that loop expansions do happen from active regions as reported in [15] using Yohkoh SXT data. However, the measured speeds were only of the order of a few to a few tens of km/s. This speed is much smaller than the typical speed of CMEs that end up as ICMEs: 934 km/s (MCs) and 772 km/s (non- MCs), which are much greater than the average speed of all CMEs [14]. In fact, it is well known that only fast and wide CMEs have enough energy to travel far into the interplanetary medium. Thus, the observed CME speeds in non-MC events are not consistent with the low speeds expected from active region loop expansion. Another report concerning polar crown filament (PCF) eruptions suggested that the associated CMEs should not be considered as CMEs, but merely coronal mass expansions similar to the slow solar wind rather than flare-associated ejections [16]. We present observations of an actual PCF eruption and the associated CME, which contradict this suggestion.

Figure 2 shows the PEA following the eruption of a PCF from near the south pole observed on 2012 March 12. The eruption occurred from the southeast limb (Figure 2a) as indicated by the EUV images obtained by the Solar Dynamics Observatory (SDO). The eruption was also observed as an elongated feature by the Extreme Ultraviolet Imager (EUVI) on board the Solar Terrestrial Relations Observatory (STEREO) as an elongated feature (Fig. 2b). In the STEREO-behind (STB) view, the eruption occurs straight to the south. In both SDO and STB images, one can see extended brightening similar to PEAs. Figure 2c shows the intensity of EUV emission (I) in the box around the PCF as shown in Fig. 2b. The increase is similar to gradual long duration flares typically observed in soft X-ray light curves. The time derivative of the EUV intensity profile (dI/dt) is roughly similar to the acceleration of the eruptive PCF and the leading edge of the CME measured in STB COR1 field of view (FOV). The acceleration of the CME peaks around 0.15 km/s$^2$, not too different from the values found in low-latitude CMEs associated with filament eruption [17]. Movies of the eruption from SDO and STEREO clearly show typical flare-like scenario with the filament eruption followed by PEA. The energy required for PEA is expected from nonthermal electrons accelerated in the reconnection region as in normal flares. The CME continued to propagate in the Solar and Heliospheric Observatory (SOHO) mission's Large Angle and Spectrometric Coronagraph (LASCO) FOV with a residual acceleration of ~9 m/s$^2$, which is also typical of CMEs accelerating in the LASCO FOV. The CME had an average speed of 638 km/s within the LASCO FOV and the final speed was about 750 km/s when the CME left the LASCO FOV (details can be found in http://cdaw.gsfc.nasa.gov/CME_list/UNIVERSAL/2012_03/univ2012_03.html). Note that these speeds are not too different from the average speed of CMEs from the disk center that became ICMEs [14]. The height-time profile of this CME is not at all like the slow solar wind and seems to be magnetically driven like any other low-latitude CME. In fact the CME acceleration and the time derivative of the EUV intensity resemble the Neupert effect well known in regular flares [18]. Thus, we conclude that even CMEs associated with PCF are not coronal mass expansions. The inference, then, is that the non-MC ICMEs are also formed similar to the MCs with the PEAs and flux ropes. Combined with the fact that a flux rope can be fit to CMEs near the Sun irrespective of the nature of their IP counterparts [19] and that CMEs becoming non-MCs tend to be deflected away from the Sun-Earth line [20], we suggest that the flux rope model is good for both MCs and non-MCs and that all CMEs contain a flux rope. These observations lend support to the possibility that the flux rope structure in non-MCs is not encountered by the observing spacecraft. This can happen when the observing spacecraft passes through the edges of the flux ropes, so the single-point in situ observations do not show the usual magnetic signatures of an MC, as explained in [4].

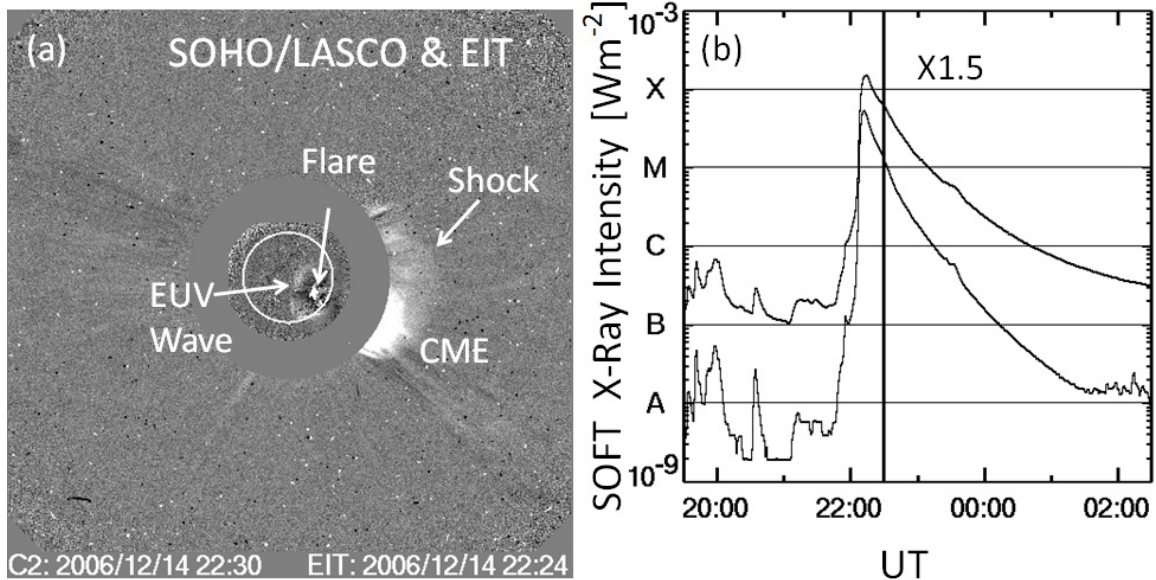

**FIGURE 3.** (a) The solar source of the 2006 December 14 CME, with the flare, EUV wave, and the shock surrounding the CME marked. (b) The GOES soft X-ray light curves showing the X1.5 flare in two energy channels (1 -8 A upper curve and 0.5 – 4 A lower curve.. The vertical line marks the time of the CME in the top left panel (22:30 UT).

## THE COMPLETE CME STRUCTURE NEAR THE SUN

The suggestion by Gold [21,22] of IP shocks driven by a magnetic structure from the Sun was soon verified by the Mariner 2 observations [23]. White-light CMEs when discovered in the early 1970s, were readily identified with the "Gold bottle" [24]. However, the shock ahead was not identified in white light, even though the existence of shocks has been inferred from type II radio burst observations [25] long before the discovery of CMEs. The shock features surrounding the bright CME material in coronagraph images were first identified only recently [26]. Recent observations suggest that the shock signature is quite robust in fast CMEs near the Sun [26-29]. The shock signature has also been reported based on EUV images and shock heating [30-31]. In fact, the standoff distance of the shock from the flux rope and the radius of curvature of the flux rope are directly measurable in coronagraphic and EUV images and can be related to the Alfvenic Mach number in the corona. With the density measurements either from the coronagraphic images or from the band splitting of type II radio bursts, one can determine the coronal magnetic field into which the CME-driven shocks propagate [29-30]. In order to complete the CME picture consistent with observations near the Sun and in the solar wind, one has to expand the picture of the three-part structure into four- or five-part structure to include the shock sheath and the shock itself. Such a picture is readily discernible from events like the one shown in Figure 3. What is shown is a complete eruption including the flare, the CME, and the CME-driven shock using a SOHO/LASCO image superposed on EUV image obtained by the Extreme Ultraviolet Imaging Telescope (EIT) on SOHO. The flare is the compact structure surrounded by the EUV wave driven by the CME. In the coronagraphic FOV, the EUV wave appears as the white-light shock. In reality, the diffuse structure surrounding the bright CME material is the compressed sheath region. The shock itself is too thin to be imaged by coronagraphs. The global picture seen in Fig. 3 is consistent with the ICME picture traditionally obtained from in situ observations: the shock, sheath and ICME. One crucial difference is the lack of magnetic properties of the flux rope in the coronagraphic images, whereas they are readily available from in situ observations. The CME core is not visible in the difference image shown in Fig.3a, because it is well below the CME material. When a spacecraft moves from right to left through the eruption, one would see a structure similar to that of shock-driving ICMEs. In situ observation of this five-part structure should be possible when the Solar Orbiter and Solar Probe Plus missions make CME measurements close to the Sun in the near future.

## SUMMARY

The standard model of CME eruptions, known as the CSHKP model seems to be readily supported by observations. One of the consequences of such a model is the formation of a post-eruption arcade and a flux rope. The post-eruption arcade (or flare loops) represents plasma heating associated with the eruption. The flux rope is expelled from the Sun as the basic structure of a CME that can be observed in the solar wind as MCs. The possibility of pressure-driven loop expansion explaining non-MCs runs into difficulties mainly because the associated CMEs are fast and wide similar to the ones associated with MCs. The post-eruption arcades associated with non-MCs are also similar to those in MC events suggesting that the basic process is identical [32]. The similarity in the charge state enhancement within MCs and non-MCs also points to the same eruption process for the two types of magnetic structures. The difference between MCs and non-MCs may arise due to the non-radial propagation of some flux ropes either due to inherent eruption direction or due to deflection by other coronal structures such as coronal holes [33] and CMEs [34]. There are also other effects such as the interchange reconnection [35] that affect the evolution of flux ropes in the interplanetary medium. The interchange reconnection is supposed to take place between closed field structures such as flux ropes and nearby open field lines [36]. Numerical and analytical modeling of this process has shown that flux ropes do erode and that 40-80% of the field lines remain closed in the inner heliosphere [37]. In bigger CMEs, ~80% of flux remains closed, which is relevant to ICMEs we are interested in. Counter-streaming signatures identified in 31 Ulysses MCs at ~5 AU indicate that the fraction of closed field range from 0 to 100%, with an average of 55% [38]. These observations are consistent with the modeling work mentioned earlier [37] when scaled to the 1 AU distance. How this flux rope erosion affects the appearance of an ICME as MC or non-MC needs further investigation. Note that the interchange reconnection is a propagation effect, whereas the flare reconnection is an effect that forms the flux rope and the charge state enhancement.

Finally, fast CMEs driving a shock are readily identified in coronagraphic images. The CSHKP model becomes complete if one includes the shock structure surrounding the flux rope in the case of fast CMEs.

## ACKNOWLEDGMENTS

The author thanks P. Mäkelä and S. Akiyama for help with the figures. Work supported by NASA LWS TR&T program.